\documentclass[%
 reprint,
 amsmath,amssymb,
 aps,
prl,
]{revtex4-2}

\usepackage{graphicx}
\usepackage{dcolumn}
\usepackage{bm}

\begin{document}

\preprint{APS/123-QED}

\title{The folding of art: avoiding one's past in finite space}

\author{Anders Levermann}
\email{anders.levermann@pik-potsdam.de}
 \homepage{http://www.pik-potsdam.de/~anders}
\affiliation{Department of Complexity Science, Potsdam Institute for Climate Impact Research, Potsdam, Germany}%
\affiliation{Lamont-Doherty Earth Observatory, Columbia University, Palisades, NY, USA.}
\affiliation{Department of Physics, Potsdam University, Potsdam, Germany.}


\date{\today}

\begin{abstract}
Through-out human history the new generations have sought to create their own artistic style while trying to avoid repeating, for example, earlier generations' music. If we assume that this search occurs in a multi-dimensional but confined space of creativity, this gives rise to highly complex dynamics. We present a very simple mathematical model with two parameters which can serve as a generic representation for {\it past avoidance} in finite space and is qualitatively distinct from earlier dynamical systems. In the presented radially confined form, the trajectory preserves its complexity while retracting to the vicinity of a hypersurface of constant radius when considered in higher dimension.

\end{abstract}

                              
\maketitle



Complex dynamical behaviour that emerges from simple rules of a small number of actors is still a non-trivial phenomenon despite it being the subject of non-linear dynamical systems theory for decades.
Complex but structured behaviour can emerge from deterministic non-linear dynamics of a single trajectory forming, for example, a strange attractor over time \cite{lorenz63,sprott14} or from spatially evolving patterns such as Laplacian growth  \cite{paterson81,barra_davidovitch01,hentschel_levermann02}  or it can emerge from unstructured stochasticity such as, for example, in the time evolution of single random walkers in diffusion-limited aggregation \cite{witten_sanders81,hastings_levitov98,davidovitch_jensen01} or simultaneously evolving fracture dynamics \cite{lemaire_levitz91,zhao_maher93,levermann_procaccia02} or self-organized criticality \cite{bak_tang87,tang_bak88,peters_hertlein02,scheinkman_woodford94}. There are many qualitatively different ways in which complexity can arise from simple rules.

Here we present a deterministic motion in $D$-dimensional space which shows highly complex behaviour through avoiding its own past and being confined to a finite space. The complexity arises from an ever expanding dynamics that is confined to finite space. The main difference to traditional dynamical systems is that it is non-local in time. We present two mathematical representations: one through a time-dependent potential landscape and the other through an integro-differential equation. Geometrically the systems exhibits the trajectory folding that is found in strange attractors and is thereby a relatively simple and pure prototype of the dynamical folding principle. Note however that the trajectory can cross itself because it tracks its history and thereby the state space is not a phase space.

As a motivation for the model we suggest the path of society in search for cultural innovation. It can also be considered a paradigmatic system for infinite economical growth on a finite planet with finite resources. The equations themselves resemble those of a relativistic description of a charged particle via the Liénard-Wiechert potential \cite{lienard1898,wiechert1901}, in which the electron is repelled by its own trajectory. But these are merely illustrations of possible applications. The main result is the complexity that arises from a spatial confinement of infinite growth.


In order to capture, for example, the path of society in an abstract creativity space we consider its trajectory ${\bf r} \in \mathbb{R}^D$ changing position along the negative gradient of a landscape $L$.
\begin{equation}
\frac{d{\bf r}}{dt} = - {\bf \nabla} L \left({\bf r} \right)
\label{equ_drdt}.
\end{equation}
The landscape $L$ is defined everywhere within a certain bounded domain $\cal B$ as the sum
\begin{equation}
L\left({\bf x}, t\right) = L_b\left({\bf x}\right) +  L_h\left({\bf x}, t\right)
\label{equ_L}.
\end{equation}
 of a {\it boundary potential}, $L_b$, and a {\it history potential}, $L_h$, with ${\bf x} \in {\cal B} \subset \mathbb{R}^D$. Here we choose the bounded domain to be the unit sphere in $D$ dimensions. The time-independent boundary potential, $L_b$, is thus defined as
\begin{equation}
L_b \left({\bf x}\right) = \left( 1 - \left| {\bf x} \right| \right)^{-1}
\label{equ_Lb}.
\end{equation}
with $|{\bf x}| \equiv \sqrt{\sum_{j=1}^D x_j^2}$ the absolute value.
The history potential, $L_h$, is defined through an ordinary differential equation via
\begin{equation}
\frac{dL_h}{dt}\left({\bf x}\right) = \left( 1 + \frac{\left| {\bf r}(t) - {\bf x} \right|}{\rho} \right)^{-1}- \frac{L_h}{\tau}
\label{equ_Lh}.
\end{equation}
with positive constants $\rho>0$ and $\tau > 0$. These two parameters have a clear interpretation. While $\tau$ controls the memory of society with respect to its history, $\rho$ defines the space in creativity space that is taken by the current generation. The assumption that these are constant is made for simplicity.
\begin{figure}
    \centering
    \includegraphics[width=.5\textwidth]{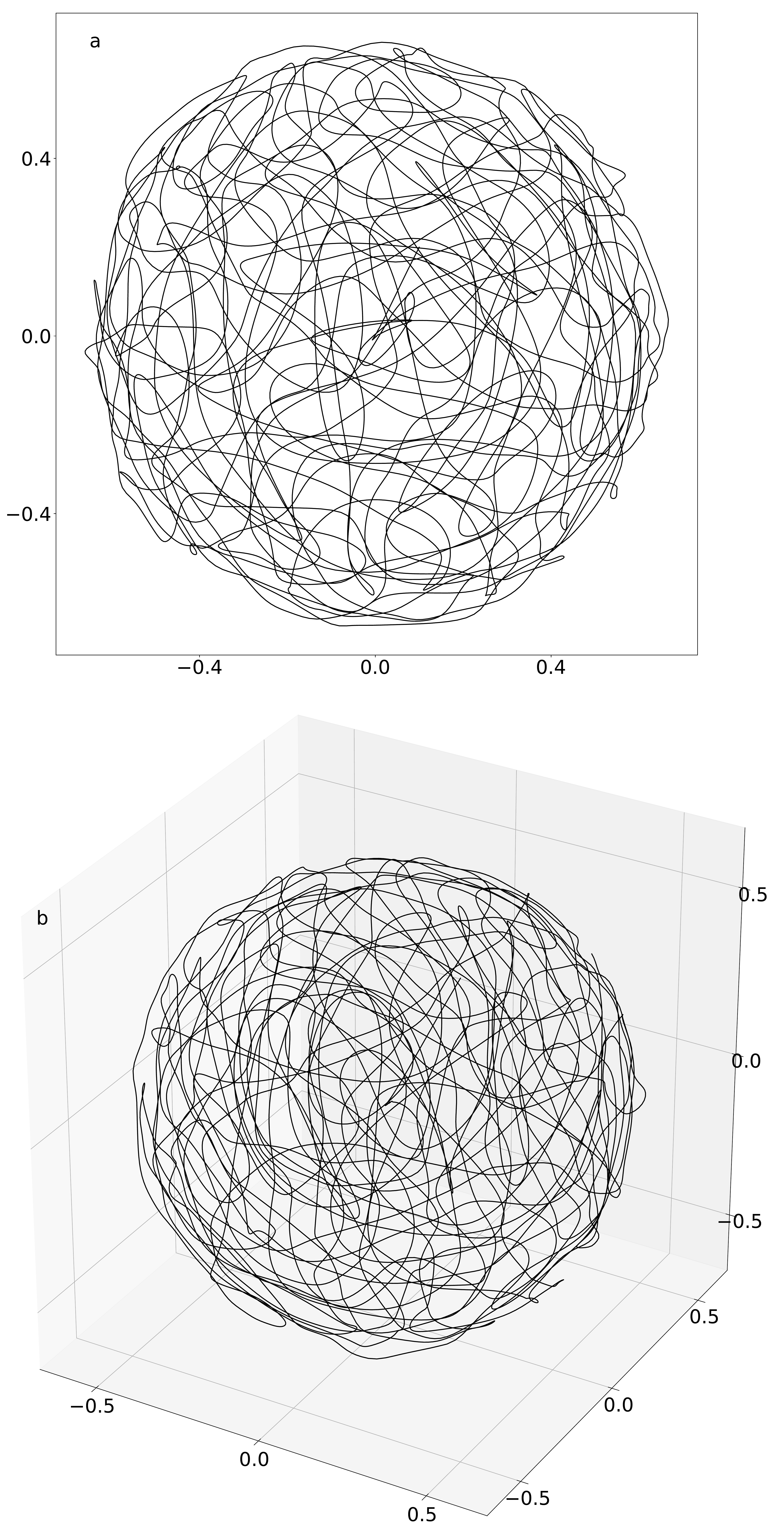}
    \caption{Generic trajectory of the dynamics (panel a)  in $D=2$ with parameters $\rho = 0.1$ and $\tau = 50$ and (panel b) in $D=3$ dimensions with $\rho = 0.1$ and $\tau = 100$. Shown is the initial time evolution up to $T=200$ for initial con}
    \label{fig_attractor}
\end{figure}
The kernel of the history potential (fig.~S1 in the supplementary material)
\begin{equation}
K\left({\bf r}(t),{\bf x}\right) = \left( 1 + \frac{\left| {\bf r}(t) - {\bf x} \right|}{\rho} \right)^{-1} = \frac{\rho}{\rho + \left| {\bf r}(t) - {\bf x} \right|}
\nonumber
\end{equation}
is one at the current position of the trajectory $K\left({\bf r}(t),{\bf x}={\bf r}(t)\right) = 1 $, and declines (quasi-hyperbolically) to $K = 0.5 $ at a distance $\rho$ from the current trajectory position ${\bf r}(t)$. Thus according to equation~(\ref{equ_Lh}) the trajectory generates an exponentially decaying tail along its path.
Thus fittingly, the set of equations can be rewritten as an integro-differential equation (see appendix\ref{app_intdiff_equ}).

If we start with a generic history landscape that is equal to zero everywhere $L_h({\bf x},t=0) = 0; \forall \; {\bf x} \in \mathbb{R}^D$, then the trajectory ${\bf r}(t)$ is fully determined by its initial position ${\bf r}(t=0) = {\bf r_0}$ which, in principle, is an additional free parameter, but the statistics of the emerging attractor does not seem to be dependent on this initial condition. A coefficient that weighs the two contributions from $L_b$ and $L_h$ can be incorporated into the choice of the radius of the sphere of the boundary potential. Here we chose the unit circle and an equal weighting of the two potentials for illustrative purposes.


Starting from any initial condition the resulting trajectories are complex and form a ball in their respective creativity space (fig. \ref{fig_attractor} and animation in supplementary information). Without proof we hypothesis that the statistics of the emerging trajectory is independent of the initial condition. There might be a specific set of initial conditions that yield regular behaviour, but placing the initial condition in the origin does yield the same complex behaviour that cannot be easily distinguished from that of fig.\ref{fig_attractor}. The trajectory exhibits strong sensitivity to the initial condition (Fig. S2), but the long-term trajectory density does not. It only depends on the two parameters, i.e. the systems memory $\tau$ and the space occupied by the current generation $\rho$. 

Due to the exponential decay of the memory the long-term average of the history potential is likely to reach a steady state given by
\begin{equation}
\left<L_h\right> (r) =\frac{\tau \rho}{2 \pi} \int_0^1  \int_0^{2\pi} \frac{p(r^\prime) {r^\prime}^{D-1} dr^\prime d\theta}{\rho + \sqrt{ r^2 + {r^\prime}^2 - 2 r r^\prime \cos(\theta) }} 
\end{equation}
where $p(r)$ is the long-term trajectory density which is given by an implicit equation
\begin{equation} \label{pot_equ}
 \frac{\tau \rho}{2 \pi} \int_0^1  \int_0^{2\pi} \frac{\partial}{\partial r} \frac{p(r^\prime) {r^\prime}^{D-1} dr^\prime d\theta}{\rho + \sqrt{ r^2 + {r^\prime}^2 - 2 r r^\prime \cos(\theta) }} =  \frac{1}{ (1 - r)^2 }.
\end{equation}
as derived in the appendix.

\begin{figure}
    \centering
    \includegraphics[width=.5\textwidth]{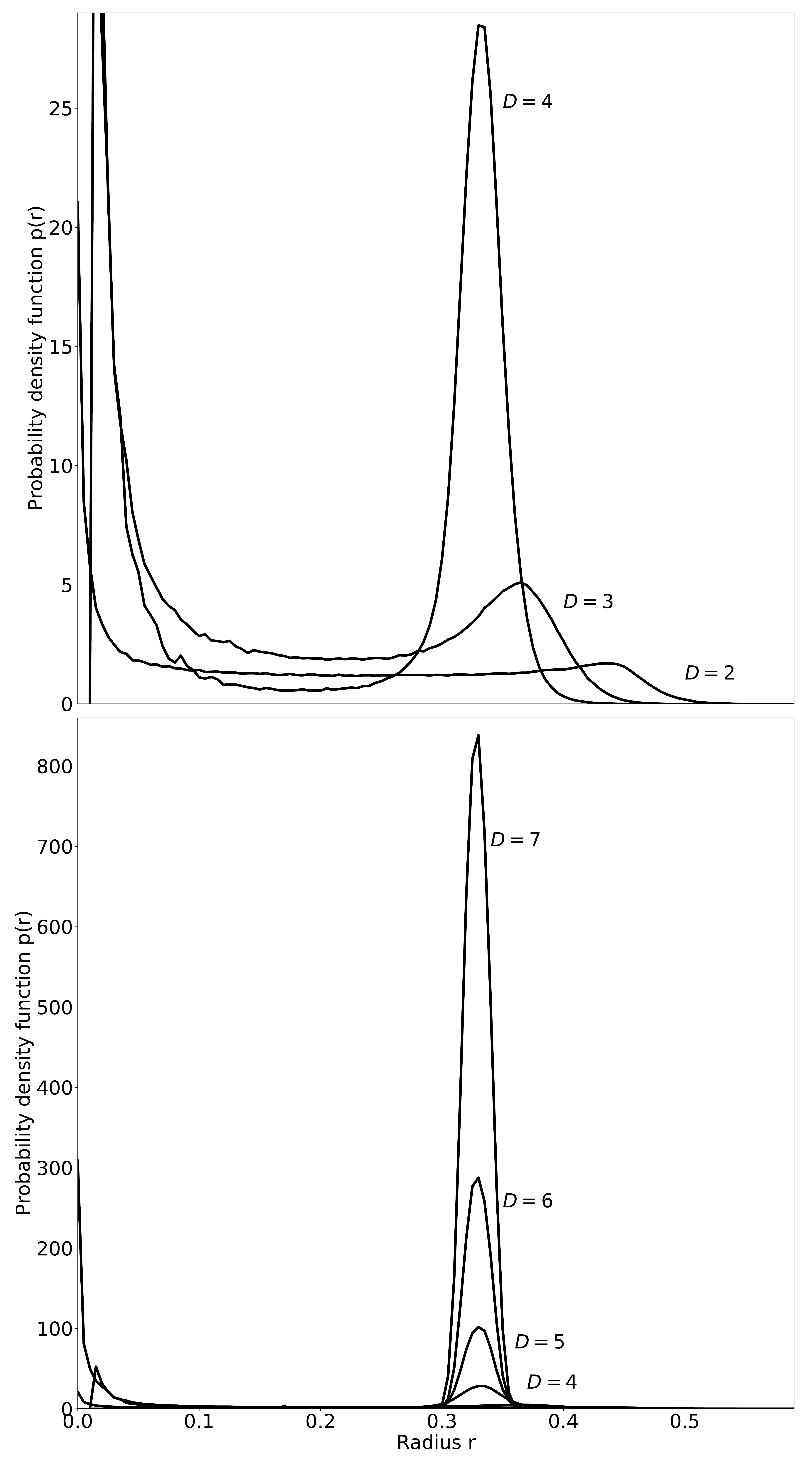}
    \caption{Density of trajectory as a function of radius $r$ in (panel a) $D=2, 3, 4$ dimensions and (panel b) $D=2,3,4,5,6,7$ dimensions for parameters $\rho = 0.1$ and $\tau=10$ which yields the convergence radius estimate of $r_m \approx 1/3$. $p(r)$ scales with $r^D$ such that $\int_0^1 dr \; r^D  p(r) = 1$}
    \label{fig_density}
\end{figure}

\begin{figure}
    \centering
    \includegraphics[width=.5\textwidth]{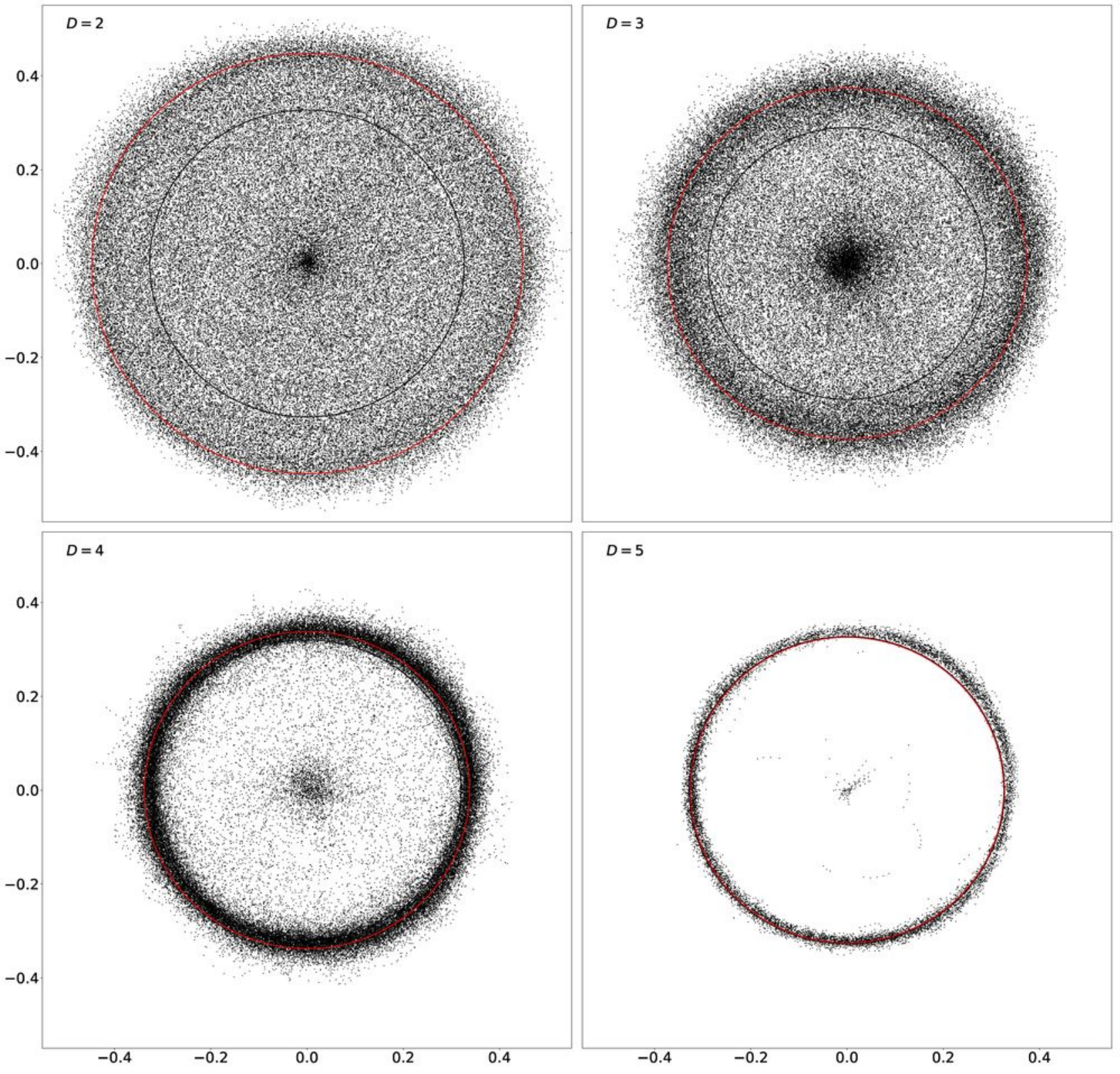}
    \caption{Projection of trajectory cloud onto plane $\cal P$ through origin in different dimensions $D$.}
    \label{fig_proj}
\end{figure}

\begin{figure}
    \centering
    \includegraphics[width=.5\textwidth]{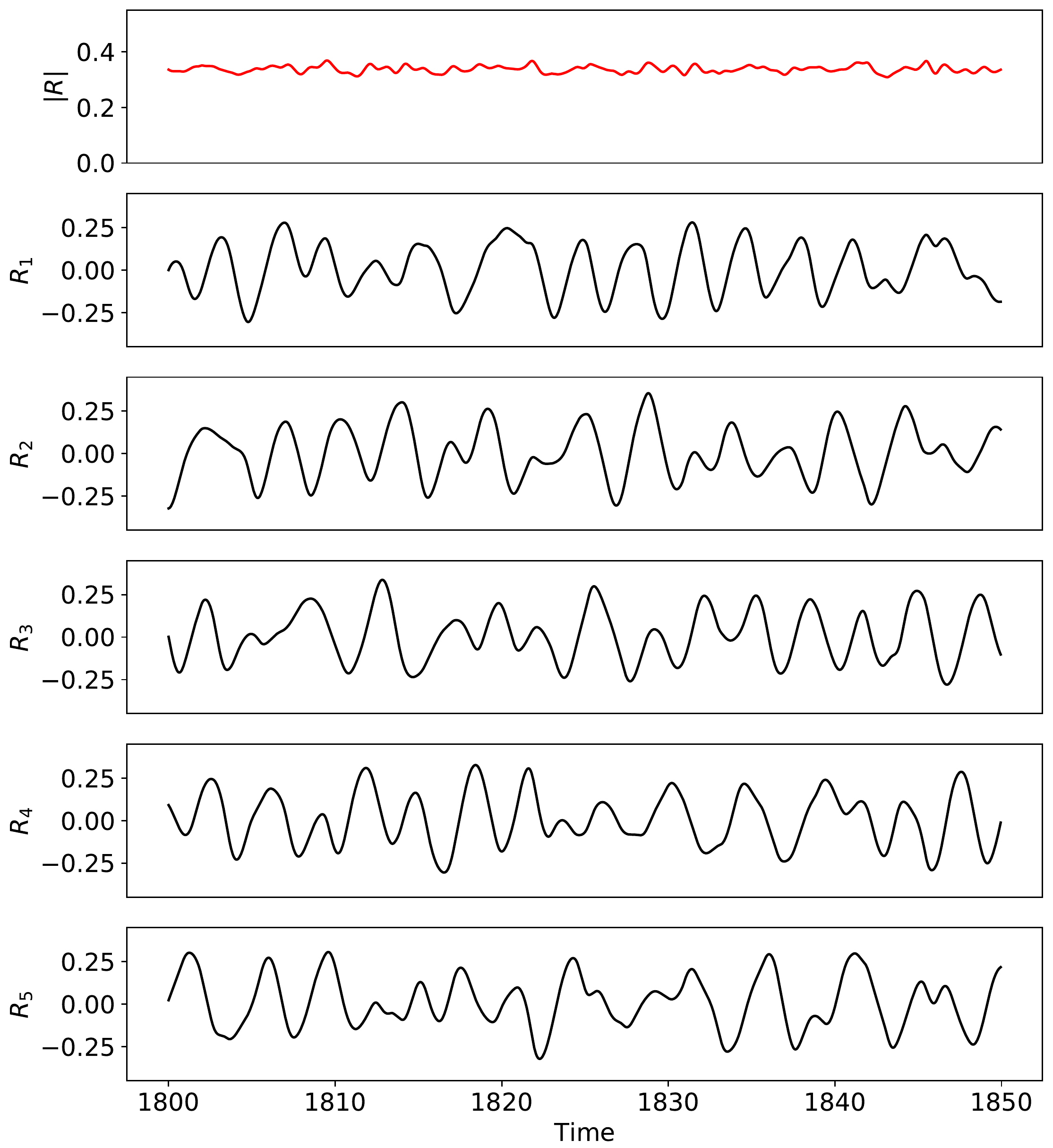}
    \caption{Example of a part of the time series of all components in $D=5$.}
    \label{fig_time}
\end{figure}

When considered in higher dimensions $D$, the dynamics retracts onto a thin shell around a hypersphere of a constant radius and some trajectory density near the origin (fig.~\ref{fig_density}). Fig.~\ref{fig_proj} shows the trajectory points that pass through a two-dimensional plane ${\cal P} = \{ x \in {\cal R}^D | x_1 = x_2 = 0\}$. This retraction towards the vicinity of a hypersurface does, however, not reflect a regularization of the dynamics. Also in higher dimensions the dynamics remains irregular while being fully deterministic. As an example fig.~\ref{fig_time} shows the time series of the five components of the Cartesian coordinates of the dynamics in $D=5$ dimensions after an arbitrarily chosen time of $t=1800$.

%
The deterministic system exhibits complex behaviour and even though the long-term radial trajectory distribution can be computed through a relatively simple equation~(\ref{pot_equ}), this does not capture the complexity of the time evolution. We hypothesize that the trajectory lies densely within the spherical shell it inhabits and is thereby ergodic. The simple set of equations~(\ref{equ_drdt}-\ref{equ_Lh}) constitutes a paradigmatic dynamics of past avoidance in finite space and a deterministic complexity dynamics that is qualitatively distinct from common deterministic chaos.

\appendix


\appendix

\section{Appendix}

\appendix

\section{Model in integro-differential form} \label{app_intdiff_equ}

The model (Eq. \ref{equ_drdt}-\ref{equ_Lh}) can be reorganized into an integro-differential equation. Since the boundary potential is time-independent its gradient at the position ${\bf r}(t)$ is simply
\begin{equation}
{\bf \nabla} L_b \left({\bf r}\right) = \left( 1 - r(t) \right)^{-2} \cdot \frac{{\bf r(t)}}{r(t)}
\label{equ_}.
\end{equation}
with $r(t) = |{\bf r(t)}|$.
Time integration and spatial differentiation of the history potential yields
\begin{equation}
{\bf \nabla} L_h \left({\bf r}(t),t\right) = \int_0^t dt' \frac{\rho   \cdot e^{-\frac{t-t'}{\tau}} }{ \left(\rho + \left| {\bf r}(t) - {\bf r}(t') \right|\right)^2}\cdot \frac{{\bf r}(t) - {\bf r}(t')}{\left| {\bf r}(t) - {\bf r}(t') \right|}
\label{equ_}.
\end{equation}
which results into the integro-differential form 
\begin{eqnarray}
\frac{d{\bf r}}{dt} (t) &=& - \left( 1 - r(t) \right)^{-2} \cdot \frac{{\bf r(t)}}{r(t)} \nonumber \\
 && - \int_0^t dt' \frac{\rho   \cdot e^{-\frac{t-t'}{\tau}} }{ \left(\rho + \left| {\bf r}(t) - {\bf r}(t') \right|\right)^2}\cdot \frac{{\bf r}(t) - {\bf r}(t')}{\left| {\bf r}(t) - {\bf r}(t') \right|}
\label{equ_}.
\end{eqnarray}
\section{Emergent steady state potential} \label{app_potential}

When taking the limit of the time-average of equ.~(\ref{equ_Lh}) the left hand side yields
\begin{equation}
\lim_{T\rightarrow \infty} \frac{1}{T} \int_0^T \frac{dL_h}{dt}\left({\bf x}\right) = \lim_{T\rightarrow \infty} \frac{1}{T} \left( L_h(T) - L_h(0) \right) = 0
 \nonumber
\end{equation}
as long as the potential remains finite within $\cal B$. The right hand side becomes
\begin{eqnarray}
\lim_{T\rightarrow \infty} \frac{1}{T} \int_0^T \left( 1 + \frac{\left| {\bf r}(t) - {\bf x} \right|}{\rho} \right)^{-1} - \lim_{T\rightarrow \infty} \frac{1}{T} \int_0^T  \frac{L_h}{\tau} \nonumber \\
=
\lim_{T\rightarrow \infty} \frac{1}{T} \int_0^T \left( 1 + \frac{\left| {\bf r}(t) - {\bf x} \right|}{\rho} \right)^{-1} -   \frac{\left<L_h\right>}{\tau}
 \nonumber
\end{eqnarray}
and thus the long-term time average of the history potential, assuming it exists, can be computed as
\begin{equation}
\left<L_h\right> = \lim_{T\rightarrow \infty} \frac{\tau}{T} \int_0^T \left( 1 + \frac{\left| {\bf r}(t) - {\bf x} \right|}{\rho} \right)^{-1}.
 \nonumber
\end{equation}
Assuming the emergent potential is radially symmetric and corresponds to a time-independent probability for the trajectory position $p({\bf r}) = p(r)$ the long-term average of the history potential can be written as
\begin{eqnarray}
&&\left<L_h\right> (r)  = \tau \cdot \int_{\cal B} d{\bf r^\prime} \frac{\rho \cdot p({\bf r^\prime})}{\rho + \left| {\bf r}(t) - {\bf r^\prime} \right|}  \nonumber \\
&=& \frac{\tau}{2\pi S_{D-2}} \int_0^1 \int_0^{2\pi} \int_{S_{D-2}}  \frac{\rho \cdot  p(r^\prime) {r^\prime}^{D-1} dr^\prime d\theta d\Omega}{\rho + \sqrt{ r^2 + {r^\prime}^2 - 2 r  r^\prime  \cos(\theta) }}  \nonumber \\
&=& \frac{\tau \rho}{2 \pi} \int_0^1  \int_0^{2\pi} \frac{p(r^\prime) {r^\prime}^{D-1} dr^\prime d\theta}{\rho + \sqrt{ r^2 + {r^\prime}^2 - 2 r  r^\prime \cos(\theta) }}  \nonumber 
 \nonumber
\end{eqnarray}
where $S_{D-2}$ is the surface of $D-2$ sphere.
Following a similar reasoning as above the long-term time average of the left hand side of equ.~(\ref{equ_drdt}) vanishes 
\begin{equation}
\lim_{T\rightarrow \infty} \frac{1}{T} \int_0^T \frac{d{\bf r}}{dt} = \lim_{T\rightarrow \infty} \frac{1}{T} \left({\bf r}(T) - {\bf r}(0) \right) = {\bf 0}
 \nonumber
\end{equation}
which yields an implicit equation for the probability density function $p(r)$
\begin{equation}
 \frac{\tau \rho}{2 \pi} \int_0^1  \int_0^{2\pi} \frac{\partial}{\partial r} \frac{p(r^\prime) {r^\prime}^{D-1} dr^\prime d\theta}{\rho + \sqrt{ r^2 + {r^\prime}^2 - 2 r r^\prime \cos(\theta) }} =  \frac{1}{ (1 - r)^2 }.
\end{equation}

\bibliography{ArtFolding.bib}

\end{document}


\preprint{APS/123-QED}

\title{Supplementary Material: \\The folding of art: avoiding one's past in finite space}

\author{Anders Levermann}
\email{anders.levermann@pik-potsdam.de}
 \homepage{http://www.pik-potsdam.de/~anders}
\affiliation{Department of Complexity Science, Potsdam Institute for Climate Impact Research, Potsdam, Germany}%
\affiliation{Lamont-Doherty Earth Observatory, Columbia University, Palisades, NY, USA.}
\affiliation{Department of Physics, Potsdam University, Potsdam, Germany.}


\date{\today}

\begin{abstract}

\end{abstract}

\keywords{Dynamical systems, Complexity, Non-linear dynamics, SOC, Chaos}
\maketitle


\section{\label{potentials}Boundary and history potentials}
\begin{figure}
    \centering
    \includegraphics[width=.5\textwidth]{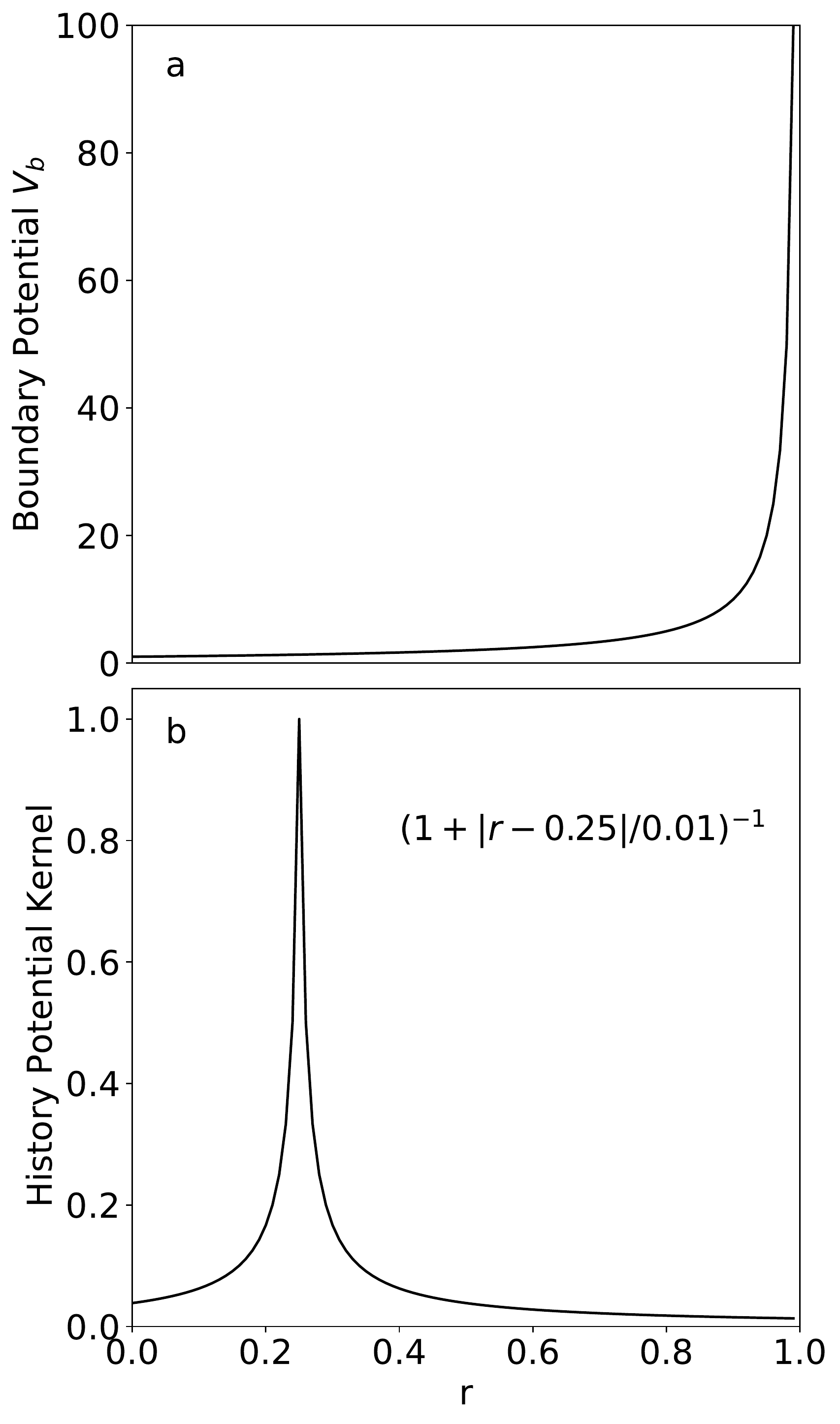}
    \caption{Illustration of the boundary potential $L_b$ (panel a) and the kernel of the history potential $L_h$ in one dimension for $\rho = 0.01$ and current position of the trajectory at $r(t) = 0.25$.}
    \label{fig_potential}
\end{figure}

The boundary potential diverges when approaching the unit circle (Fig~\ref{fig_potential}a). The history potential is generated through a kernel
\begin{equation}
K\left({\bf r}(t),{\bf x}\right) = \left( 1 + \frac{\left| {\bf r}(t) - {\bf x} \right|}{\rho} \right)^{-1} = \frac{\rho}{\rho + \left| {\bf r}(t) - {\bf x} \right|}
\label{equ_Lh}.
\end{equation}
At the current position of the trajectory the kernel is one, $K\left({\bf r}(t),{\bf x}={\bf r}(t)\right) = 1 $, and at a distance of $\rho$ from the trajectory position ${\bf r}(t)$ it is $K = 0.5 $. Thus according to equation~(\ref{equ_Lh}) the trajectory generates a tail along its path  (Fig~\ref{fig_potential}b).

\section{\label{sensitivity}Sensitivity to initial conditions}
\begin{figure}
    \centering
    \includegraphics[width=.5\textwidth]{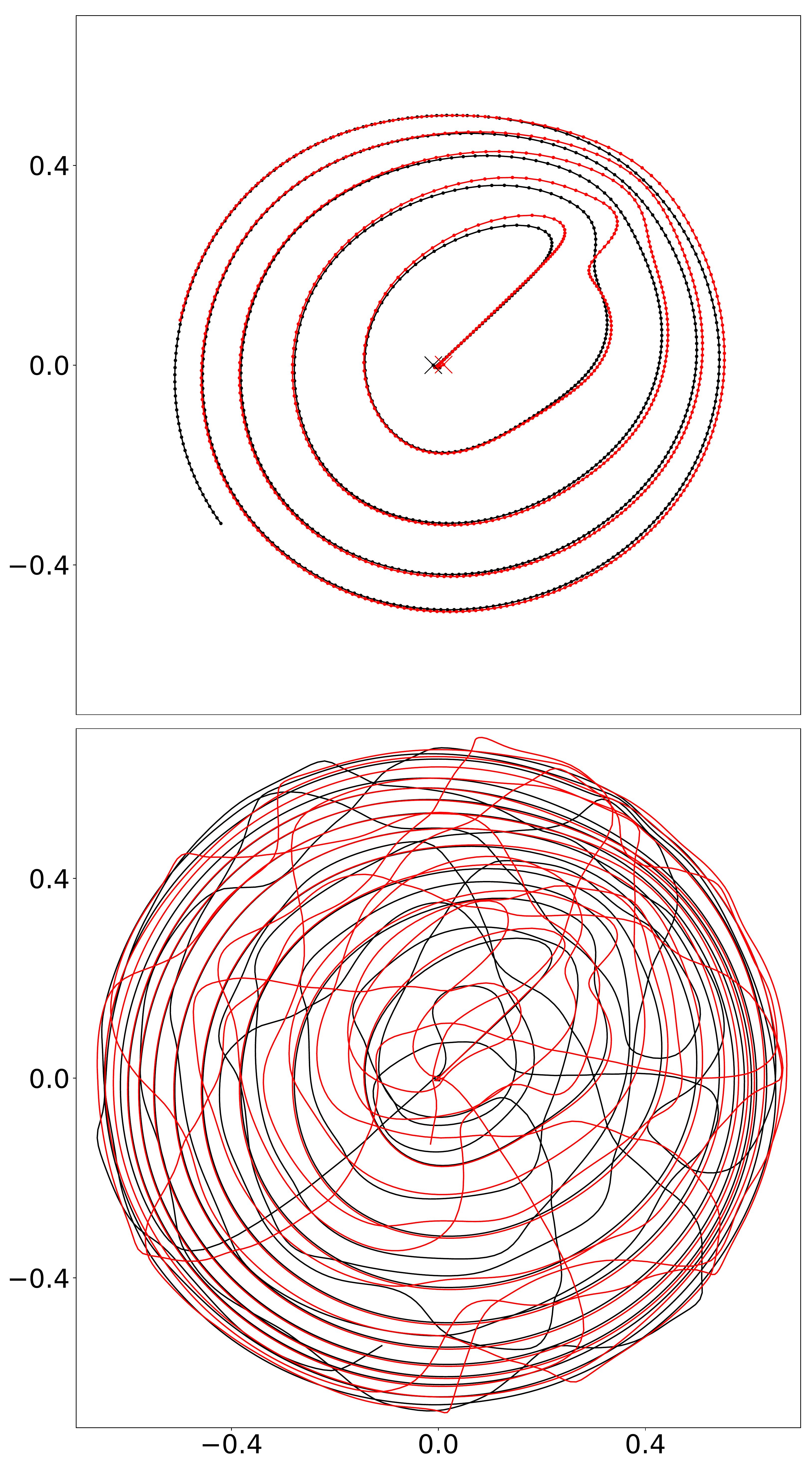}
    \caption{Two trajectories with the same parameters $\rho=0.1$ $\tau=50$ in two dimensions with initial conditions $r_0=(-0.01,0)$ (black) and $r_0=(0.01,0.001)$ (red). For the initial time period up to $T=20$ (panel a) and up to $T=100$ (panel b). The trajectories stay close to each other with some temporary deviations for some time, until the confinement forces them to fold back into the interior where the small differences yield stronger and stronger deviations between the two temporal evolutions.}
    \label{fig_init}
\end{figure}

The dynamics is deterministic but sensitive to changes in initial conditions (Fig~\ref{fig_init}).
